\documentclass[manuscript]{aastex61}

\usepackage{url}
\usepackage{tabularx}
\usepackage{amsmath}
\usepackage{rotating}
\shorttitle{LOS Characteristics and Flares}
\shortauthors{Sadykov and Kosovichev}

\begin{document}
	
\title{Relationships Between Characteristics of the Line-of-Sight Magnetic Field and Solar Flare Forecasts}

\correspondingauthor{Viacheslav M Sadykov}
\email{vsadykov@njit.edu}

\author{Viacheslav M Sadykov}
\affiliation{Center for Computational Heliophysics, New Jersey Institute of Technology, Newark, NJ 07102, USA}
\affil{Department of Physics, New Jersey Institute of Technology, Newark, NJ 07102, USA}

\author{Alexander G Kosovichev}
\affiliation{Center for Computational Heliophysics, New Jersey Institute of Technology, Newark, NJ 07102, USA}
\affiliation{Department of Physics, New Jersey Institute of Technology, Newark, NJ 07102, USA}
\affiliation{NASA Ames Research Center, Moffett Field, CA 94035, USA}

\begin{abstract}
	We analyze the relationship between the flare X-ray peak flux, and characteristics of the Polarity Inversion Line (PIL) and Active Regions (AR), derived from line-of-sight (LOS) magnetograms. The PIL detection algorithm based on a magnetogram segmentation procedure is applied for each AR with 1 hour cadence. The PIL and AR characteristics are associated with the AR flare history and divided into flaring and non-flaring cases. Effectiveness of the derived characteristics for flare forecasting is determined by the number of non-flaring cases separated from flaring cases by a certain threshold, and by their Fisher ranking score. The Support Vector Machine (SVM) classifier trained only on the PIL characteristics is used for the flare prediction. We have obtained the following results: (1) the PIL characteristics are more effective than global characteristics of ARs, (2) the highest True Skill Statistics (TSS) values of 0.76$\pm$0.03 for $\geq$M1.0 flares and 0.84$\pm$0.07 for $\geq$X1.0 flares are obtained using the ``Sigmoid'' SVM kernel, (3) the TSS scores obtained using only the LOS magnetograms are slightly lower than the scores obtained using vector magnetograms, but significantly better than current expert-based predictions, (4) for prediction of $\geq$M1.0 class flares 74.4\% of all cases, and 91.2\% for $\geq$X1.0 class, can be pre-classified as negative with no significant effect on the results, (5) the inclusion of global AR characteristics does not improve the forecast. The study confirms the unique role of the PIL region characteristics in the flare initiation process, and demonstrate possibilities of flare forecasting using only the line-of-sight magnetograms.
\end{abstract}

\keywords{methods: statistical~--- Sun: activity~--- Sun: flares~--- Sun: magnetic fields}

\section{Introduction}
\label{section1}

	Usually lasting from several minutes to several hours, solar flares can release more than $10^{32}$\,erg of energy, and cause harmful effects to the terrestrial environment. The only possible source to accumulate such large amounts of energy is magnetic field of active regions. \citet{Emslie12} demonstrated for a sample of 38 flares that the free (non-potential) energy of magnetic field was sufficient to explain the flare energy release including Coronal Mass Ejections (CMEs), energetic particles, and hot plasma emission and dynamics. For understanding the flare physical mechanism and developing flare prediction methods it is important to find critical magnetic field characteristics that are linked to the flare initiation and strength.
	
	There have been two types of such study. The first approach is to focus on global characteristics of active regions, and the second approach is to search for local critical properties of magnetic fields. For instance, in the first type studies, \citet{Mandage16} demonstrated a difference between the magnetic field power spectrum slopes of flaring and non-flaring active regions. \citet{Korsos14} found several promising preflare signatures using the SOHO/MDI-Debrecen Data sunspot catalog. \citet{Korsos15} introduced the weighted horizontal magnetic gradient, $WG_{M}$, which allowed them to predict the onset time for $\geq$M5.0 class flares, and conclude whether or not a flare is likely be followed by another event in the next 18 hours. The daily averages of $WG_{M}$ together with a separation parameter $S_{l-f}$ of magnetic polarities were used by \citet{Korsos16} to obtain some conditional probabilities of flare and CME characteristics. \citet{Bobra15,Bobra16,Nishizuka16,Liu17} have used vector magnetograms from the Space-weather HMI Active Region Patches (SHARP) and applied machine-learning techniques (Support Vector Machine, Random Forest, and Nearest-Neighbor classifiers) for flare and CME predictions. Also, a recent study of \citet{Raboonik17} used the Zerneke moments as characteristics of the active region magnetic field for flare prediction.
	
	Many observational studies of the second type found that the magnetic field Polarity Inversion Line (PIL) in regions of strong field plays an important role in the flare activity \citep[e.g.][]{Severny64,Hagyard90,Wang94,Falconer97,Kosovichev01,Jing06,Schrijver07,Kumar15,Barnes16,Schrijver16,Sharykin16,Toriumi17,Bamba17,Zimovets17}. \citet{Kusano12} demonstrated from three-dimensional magnetohydrodynamic simulations that flare eruptions can be initiated by emergence of certain small magnetic structures near PIL, as evident from observations. \citet{Toriumi13,Toriumi14} pointed out an important role of highly-sheared magnetic field in the vicinity of PILs in the flare development process. \cite{Guennou17} found from simulations that the PIL parameters measuring the total non-potentiality of active regions present a significant ability to distinguish between eruptive and non-eruptive cases. From magnetograms one can extract several descriptors representing the local field in the PIL vicinity. For example, \citet{Falconer03} showed that the length of the PIL with a strong field gradient and sheared transverse field correlates with the CME and flare productivity. \citet{Mason10} introduced the Gradient-Weighted PIL length as a characteristic for solar flare forecasts. \citet{Falconer11,Falconer12,Falconer14} found that this characteristic is a good proxy for the free magnetic energy. \citet{Leka03a,Leka03b,Leka07} suggested to use a shear angle between the observed and reconstructed magnetic fields. \citet{Chernyshov11} used the PIL length, the area of strong magnetic field in the PIL vicinity, and the total flux in this area, as well as the rates of change of these characteristics.
	
	In this paper, we perform a critical analysis of various line-of-sight (LOS) magnetic field characteristics (derived for the entire active region and for the PIL vicinity), their relationship to the flaring activity, and importance for flare forecast. Such analysis based on the LOS magnetograms is important because these observations can be performed more easily and accurately than the full vector magnetic field measurements in near-real time by various space-based and ground-based observatories. In Section~\ref{section2}, we describe automatic procedures for identification of PIL, calculation of various magnetic field characteristics, association of the derived characteristics with flare events, and construction of ``train'' and ``test'' data sets. In Section~\ref{section3}, we estimate the effectiveness in the separation of flaring and non-flaring cases for different LOS characteristics. Section~\ref{section4} describes the application of Support Vector Machine (SVM) classifier for prediction of M- and X-class flares. The results are summarizes in Section~\ref{section5}. The comparison with previous results, expert-based scores and following conclusion are presented in Section~\ref{section6}.

\section{Data Preparation}
\label{section2}

	\subsection{Magnetogram Segmentation}
	
		For analysis we used the Line-of-Sight (LOS) magnetograms of Active Regions (AR), obtained by the Helioseismic and Magnetic Imager onboard the Solar Dynamics Observatory \citep[SDO/HMI,][]{Scherrer12}. The active region data were represented in the form of $30^{o}\times{}30^{o}$ data cubes with 1\,h cadence, remapped onto the heliographic coordinates using the Postel's projection, and tracked with the solar differential rotation during the whole passage of active regions on the solar disk, employing the standard SDO software. To avoid projection effects, following~\citet{Bobra15} we consider ARs only when they are located within $\pm$68$\deg$ from the disk center.
		
\begin{figure}[p]
	\centering
	\includegraphics[width=\linewidth]{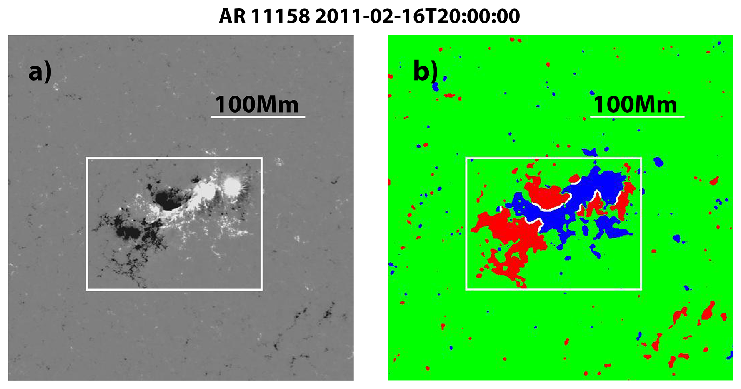}
	\caption{Illustration of the PIL automatic identification procedure: a) The magnetogram of AR 11158 obtained by the SDO/HMI at 2011-02-16 20:00:00\,UT. b) The magnetogram segmentation and identification of PIL: red, green, and blue areas correspond to negative, neutral and positive segments. The PIL identified by the algorithm described in Appendix A is shown by white curves.}
	\label{figure1_segmentation}
\end{figure}

		By definition, the Polarity Inversion Line (PIL) is the line where the LOS magnetic field changes its sign. For the automatic robust detection of the PIL of strong fields in active regions we use the algorithm initially introduced by \citet{Chernyshov11} and \citet{Laptev11}. This algorithm is based on a magnetogram segmentation process formulated as an optimization task. The goal is to divide the magnetogram into regions with strong positive field (``positive'' segments), strong negative field (``negative'' segments), or weak field (``neutral'' segments). We describe the algorithm in detail in Appendix~\ref{AppendixA}. An example of the segmentation and PIL detection for AR 11158 is illustrated in Fig.~\ref{figure1_segmentation}.
		
		To isolate the active region area, we use the following two algorithms. The first one is based on the segmentation result: we apply one morphological dilation (inclusion of neighboring pixels) to the positive/negative segments (see Appendix~\ref{AppendixA}), combine them, choose the largest segment containing the active region center, and determine the minimum bounding box around it. The second algorithm is implemented following the procedure of~\citet{Stenflo12}. The magnetogram is smoothed, and for each strong magnetic field island the bounding box with a margin of fixed width (18$^{\prime{}\prime{}}$) on all sides is defined. Then, the intersecting bounding boxes are replaced by a larger bounding box. The solution represents the largest bounding box intersecting the center of the data cube (the center of AR). We have found that by applying both algorithms and selecting the smallest bounding box almost all ARs can be effectively separated from their neighbors. The bounding box extracted for AR 11158 is presented in Fig.~\ref{figure1_segmentation}.

	\subsection{Derivation of PIL and AR Characteristics}
	
		After performing the segmentation and bounding procedures, we calculate the following descriptors (characteristics) using the derived PIL and the tracked and remapped magnetogram:
		
		\begin{enumerate}
			\item The PIL length defined as the number of pixels occupied by the PIL.
			\item The PIL area obtained after 10 morphological dilations of the PIL.
			\item The unsigned magnetic flux in the PIL area.
			\item The unsigned horizontal gradient in the PIL area defined as the sum of $\nabla{}_{h}B_{z} = \sqrt{\left(\dfrac{\partial B_{z}}{\partial x}\right)^{2} + \left(\dfrac{\partial B_{z}}{\partial y}\right)^{2}}$ over the PIL area pixels.
			\item The maximum gradient of the LOS magnetic field across the PIL.
			\item The gradient-weighted PIL length \citep{Mason10} calculated as the sum of the PIL pixels multiplied by the unsigned horizontal gradient in each pixel.
			\item The R-value \citep{Schrijver07} representing the unsigned magnetic flux weighted with the inverse distance from the PIL.
		\end{enumerate}
		
		Also, we calculate the following characteristics of the entire AR (``global'' characteristics):
		
		\begin{enumerate}
			\setcounter{enumi}{7}
			\item The AR area defined as the total area of the positive and negative segments.
			\item The unsigned magnetic flux in the AR area.
			\item The maximum strength of magnetic field in AR.
			\item The unsigned horizontal gradient in the AR area.
		\end{enumerate}

	\subsection{Definition of Positive and Negative Classes, and Construction of ``Train'' and ``Test'' Data Sets}
	
		The next important step is to associate the magnetic field characteristics derived for each AR with the flare events detected by the GOES satellite. Following \citet{Nishizuka16}, we classify a set of magnetic field characteristics as a ``positive'' case if a $\geq$M1.0 flare occurred in the corresponding AR within 24\,h after the last field measurement. This means that for each flare there can be 24 positive cases (sets of measured LOS magnetic field characteristics) or less. For the period from April, 2010 to June, 2016, 521 M-class and 31 X-class flares were associated with at least one positive case.
			
		\cite{Ahmed13} introduced two ways to determine the negative cases, described by so-called ``operational'' and ``segmented'' associations of active region characteristics and flares. According to the operational association, the negative cases are defined to be exactly opposite to the positive cases, i.e. are assigned if there was no flare of $\geq$M1.0 X-ray class within 24\,h after the magnetic field measurement. For the segmented association, the case is defined as negative if no flares occurred 48\,h before and after the case time moment. In the following we will use the operational association for the ``test'' subset while keeping the segmented association for the ``train'' subset. The segmented association better separates the positive and negative cases (by neglecting negative cases occurring very close to the flare time), while the operational association is needed for real-time predictions. The same procedure was applied also for $\geq$X1.0 class flares.
			
		For the operational-type real-time flare forecasts, the classifier is defined for future cases based on the previously observed classified cases. To simulate the real-time operational forecast, we constructed the ``train'' and ``test'' datasets to be sequential in time. We assign all the cases belonging to ARs with the NOAA numbers 11059-12158 to the ``train'' data set, and AR 12159-12559 to the ``test'' data set. The ratio of the ``train'' and ``test'' datasets is approximately 70\% to 30\% \citep[following][]{Bobra15,Nishizuka16}. We also assume that we have just one attempt to classify a ``test'' dataset for prediction of $\geq$M1.0 or $\geq$X1.0 flares, which means that the classifier tuning should be done on the ``train'' dataset only.

\section{Effectiveness of Characteristics}
\label{section3}
	
\begin{figure}[p]
	\centering
	\includegraphics[width=.49\linewidth]{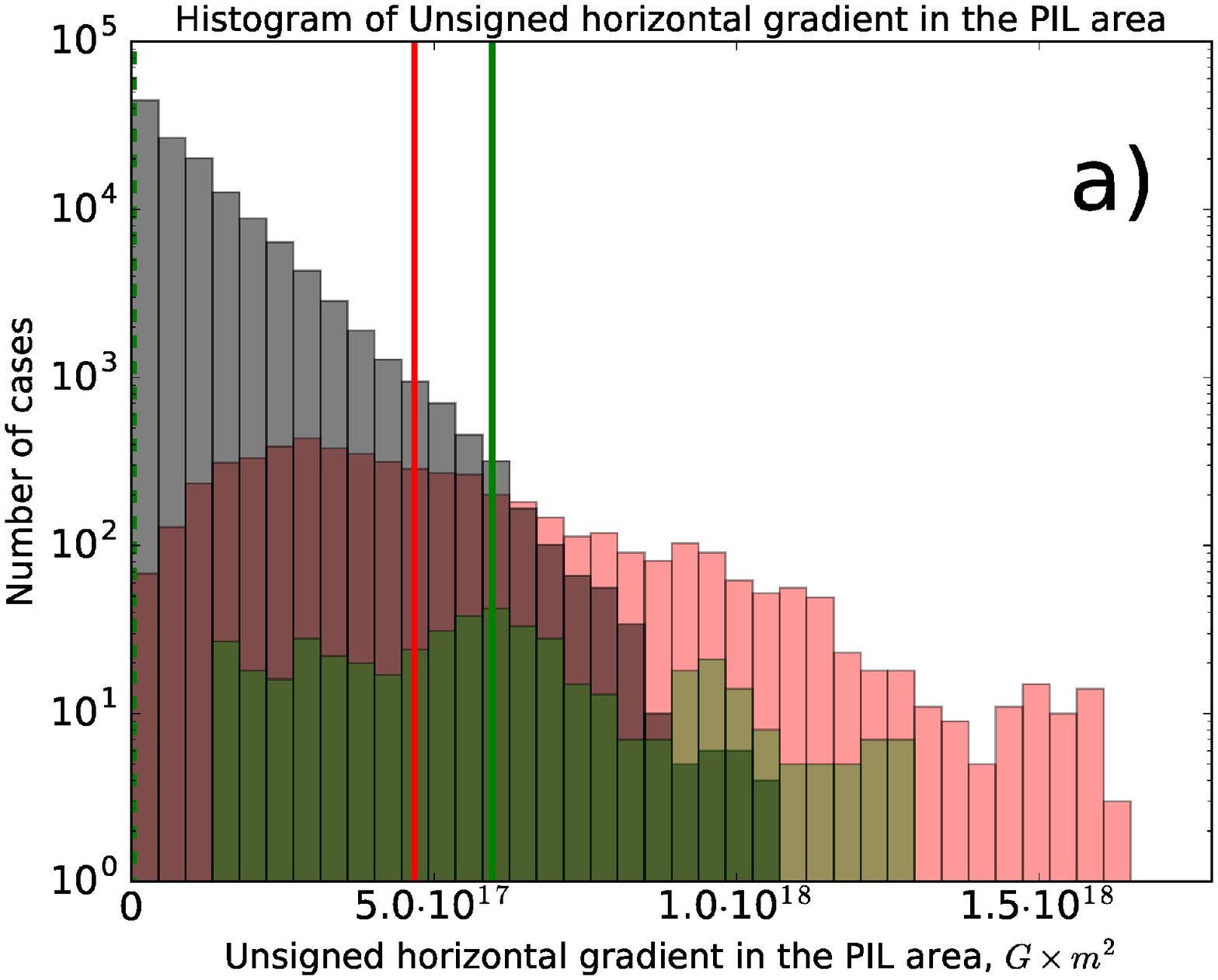}
	\includegraphics[width=.49\linewidth]{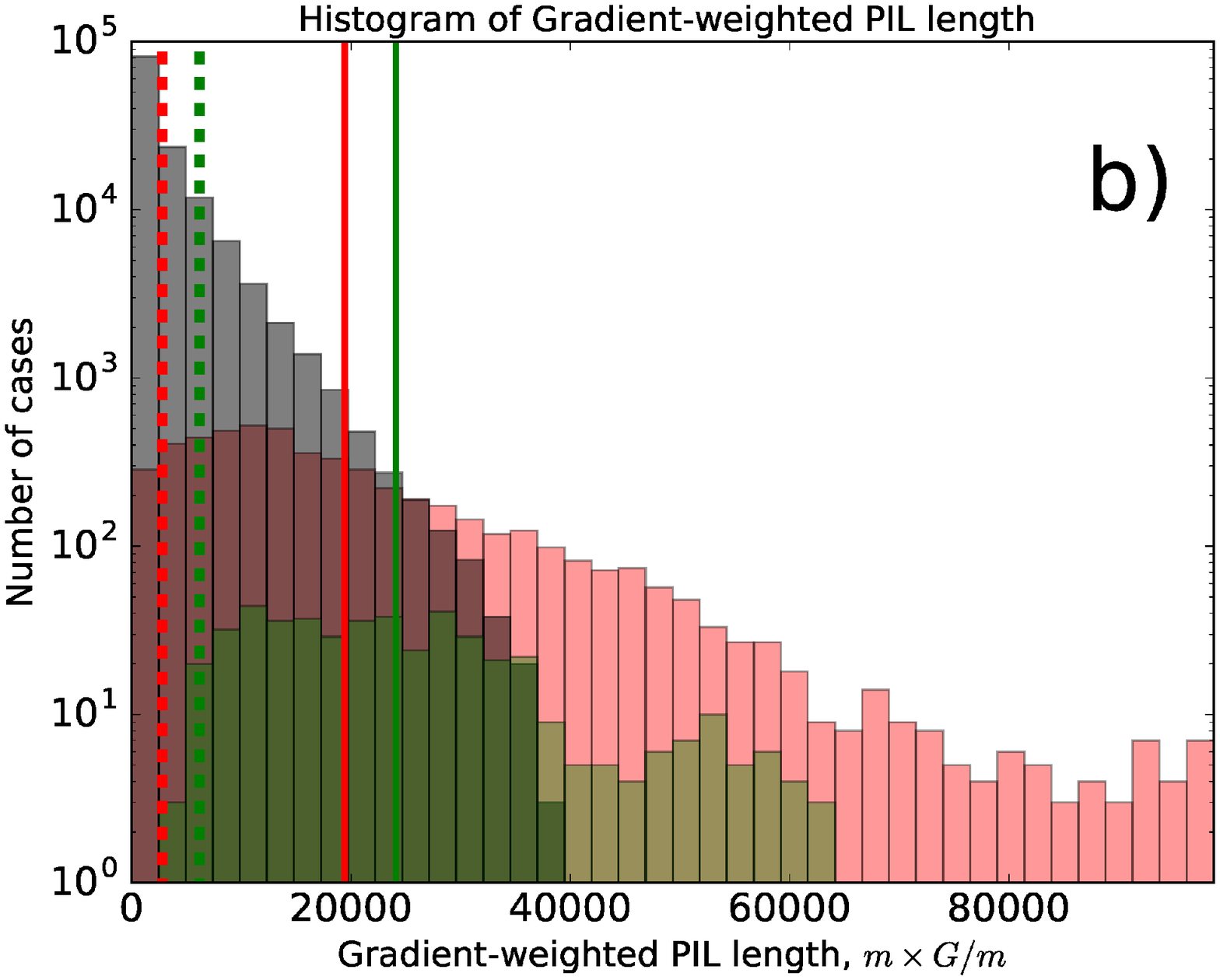} \\
	\includegraphics[width=.49\linewidth]{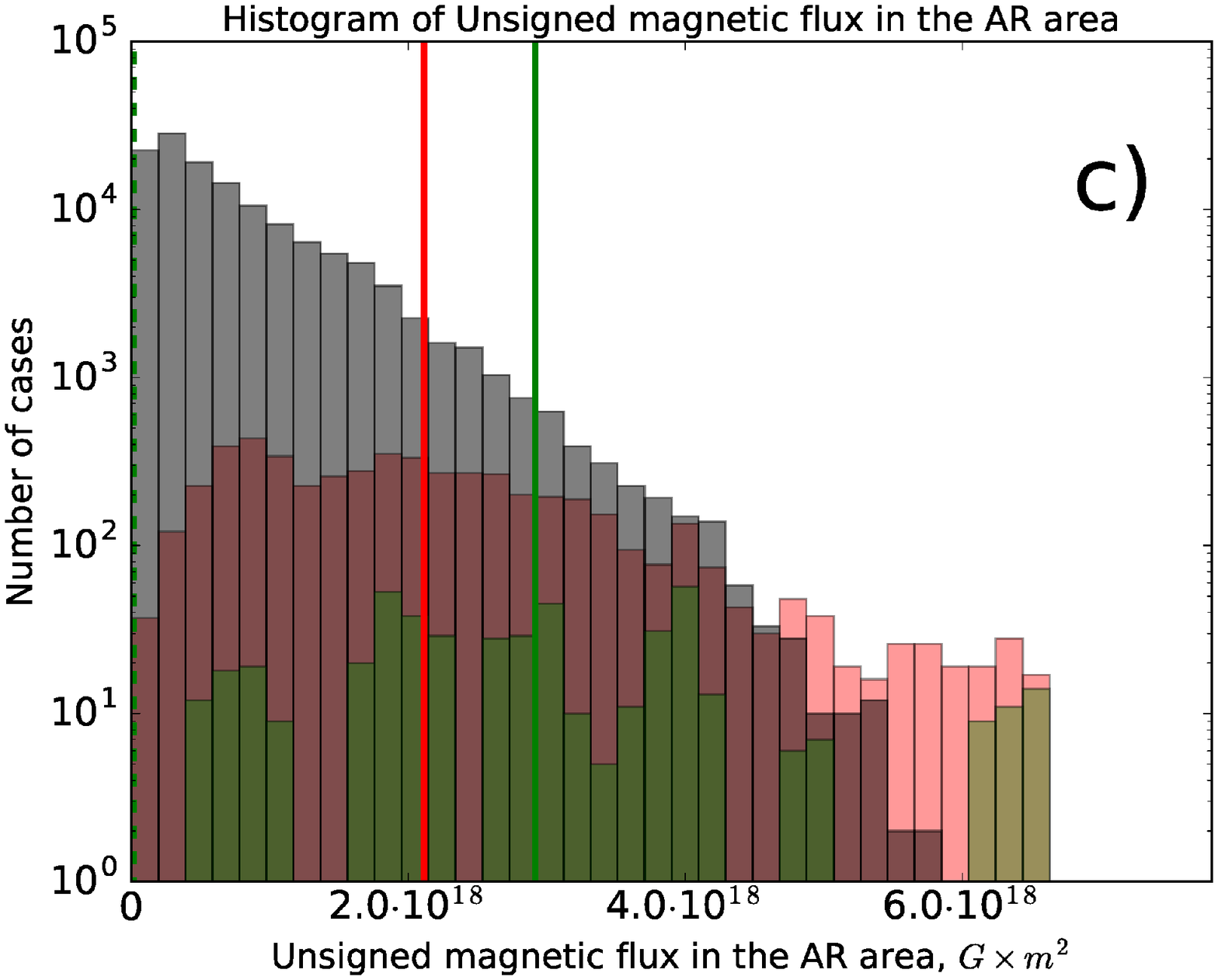}
	\includegraphics[width=.49\linewidth]{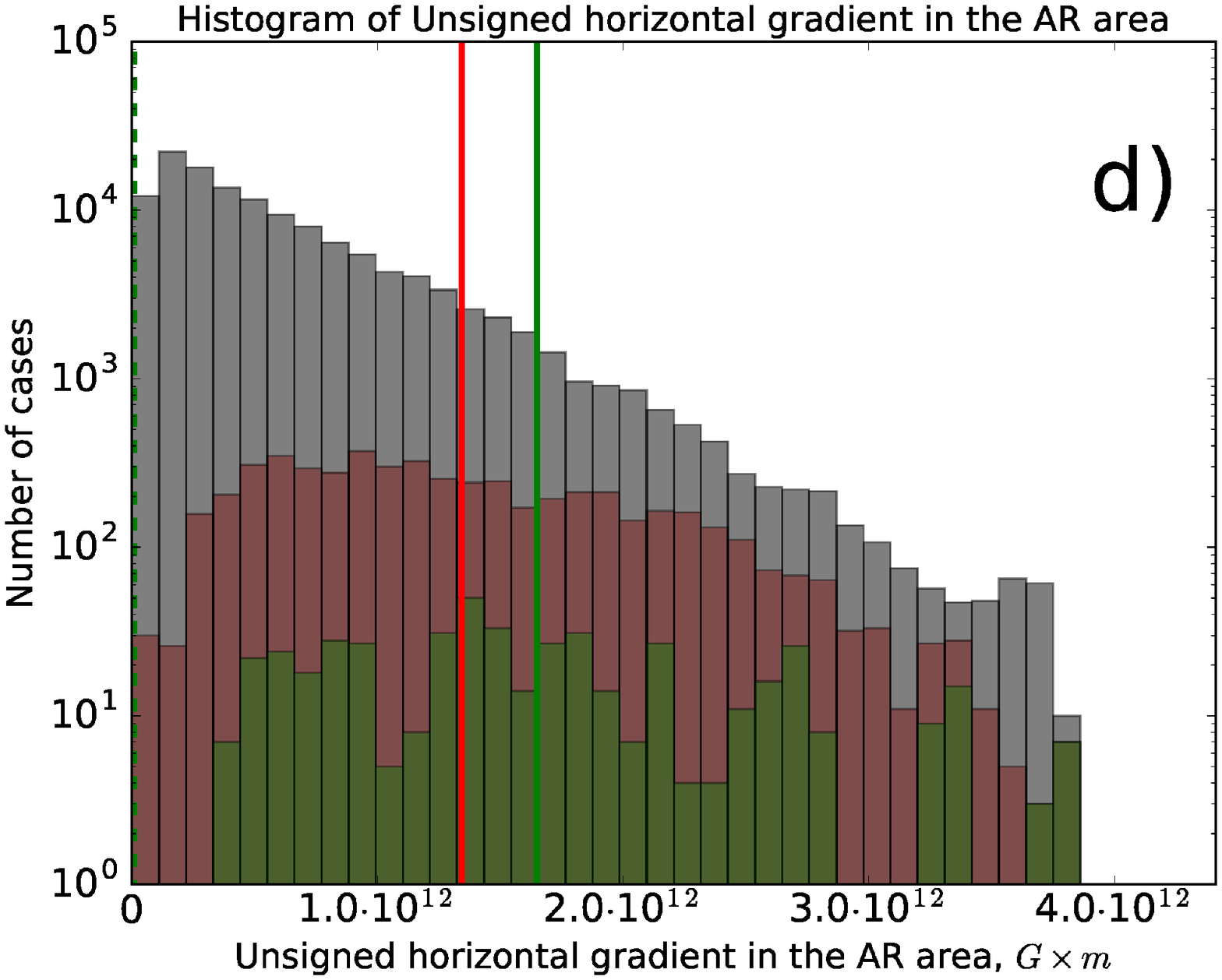}
	\caption{1D-histograms of a) unsigned magnetic flux in the PIL area; b) gradient-weighted PIL length; c) unsigned magnetic flux in the AR area; d) unsigned horizontal gradient in the AR area. The negative cases are shown in grey, the positive $\geq$M1.0 class cases are shown in red, and the positive $\geq$X1.0 class cases are shown in green. The darker areas represent the intersections of the histograms. The red and green solid lines represent the average values of the positive $\geq$M1.0 and $\geq$X1.0 cases, the corresponding dashed lines show the thresholds corresponding to 5\% of positive cases.}
	\label{figure2_general}
\end{figure}

	In this Section, we analyze the effectiveness of the derived magnetic field characteristics to separate the positive and negative (flaring and non-flaring) cases. One of the simplest ways to illustrate the separation ability of magnetic field characteristics is to construct combined histograms for positive and negative cases. The examples of such histograms are presented in Fig.~\ref{figure2_general}. The upper two panels correspond to two PIL characteristics: the unsigned magnetic flux in the PIL area and the gradient-weighted PIL length; and the lower two panels correspond to two AR characteristics: the unsigned magnetic flux in the AR area and the unsigned horizontal gradient in the AR area.
		
	One can notice that for the PIL characteristics there are more flaring than non-flaring cases in the tails of the histograms (light color areas). We found such situation for all PIL characteristics that we computed. For the global AR characteristics, we found a slight dominance of positive cases in the distribution tail only for the unsigned magnetic flux, and did not observe it for other three characteristics.
		
	There is one common feature in the histograms. The positive cases occur only if the characteristics reach some critical (threshold) value. For some LOS characteristics the existence of the critical values is more prominent in the normal-scaled histogram, but for others in the logarithmic-scaled histogram. This feature is used to simplify the classification (prediction) problem by reduction of the amounts of data considered for the classification. The red dashed (for $\geq$M1.0 flares) and green dashed (for $\geq$X1.0 flares) lines in Fig.~\ref{figure2_general} represent the threshold values, above which 95\% of positive cases are observed. Note that the threshold values are determined using the ``train'' data set. At the same time, the mean values of the positive cases are shown by solid lines of the same color. The threshold and mean values for the positive cases, as well as the mean value for the negative cases, are summarized in Table~\ref{table1_parameters}.
	
	There are many ways to quantitatively determine which characteristics are most effective for a classification problem. The inclusion of characteristics that are not discriminative leads to a high computational cost without improvement of the result, and may even decrease the performance of the SVM \citep{Bobra15}. \citet{Breiman01} proposed to evaluate feature importance by using the Random Forest classification, which was also used by \citet{Nishizuka16}. \citet{Al-Ghraibah15} employed the univariate True Skill Statistics (TSS) score as a measure of feature importance. \citet{Ahmed13} used the Correlation-Based Feature-Selection (CFS) and Minimum Redundancy Maximum Relevance (MRMR) methods. \citet{Leka03b} suggested the Mahalanobis distance between classes and Hotelling's T$^2$-test to measure statistical differences between flaring and non-flaring cases. \citet{Bobra15} calculated the Fisher Ranking score (or F-score) as a measure of a univariate effectiveness of the separation ability.
		
	In this work, we calculated two simple univariate scores for the obtained magnetic field characteristics. Firstly, for each characteristic we derived the threshold separating 5\% of the positive cases. As seen from Table~\ref{table1_parameters}, these threshold values (for both $\geq$M1.0 and $\geq$X1.0 flares) are comparable or even greater than the mean values for the negative cases for most characteristics. Thus, the fraction of negative cases which could be cut off by this threshold is used as a measure of effectiveness of characteristics in separating the ``train'' and ``test'' data sets. Secondly, we calculate the Fisher ranking score \citep[or F-score,][]{Bobra15,Chang08}:
		
	\begin{center}
		$ F(i) = \dfrac{(\bar{x}^{+}_{i} - \bar{x}_{i})^{2} + (\bar{x}^{-}_{i} - \bar{x}_{i})^{2}}
		{\dfrac{1}{n^{+}-1}\sum\limits_{k=1}^{n^{+}}(x^{+}_{k,i} - \bar{x}^{+}_{i})^{2} + \dfrac{1}{n^{-}-1}\sum\limits_{k=1}^{n^{-}}(x^{-}_{k,i} - \bar{x}^{-}_{i})^{2}},$
	\end{center}
		
	where $\bar{x}_{i}$ is the mean value of characteristic $i$; $\bar{x}^{+}_{i}$ and $\bar{x}^{-}_{i}$ are the mean values of characteristic $i$ for the positive and negative cases; and $n^{+}$ and $n^{-}$ are the total numbers of the positive and negative cases. We calculated the F-score for all the characteristics for the train dataset. Sometimes, the F-score is higher if calculated for the logarithms of the parameters. Therefore, we also calculated the F-scores of decimal logarithms of each parameter and used it if the score was higher than the one for the normal-scaled characteristic.
		
	The results for both estimates of effectiveness are combined and summarized in Tables \ref{table2_Mimportance} and~\ref{table3_Ximportance} for the $\geq$M1.0 and $\geq$X1.0 class flares respectively. The cases for which the logarithmic scale was used in the F-score calculation are labeled as (log) in Tables~\ref{table2_Mimportance} and~\ref{table3_Ximportance}. The SVM training and testing were also done in the logarithmic scale for such parameters. One can notice from Tables~\ref{table2_Mimportance}~and~\ref{table3_Ximportance} that for every considered univariate test the PIL characteristics have higher scores than the global AR parameters.

\section{Methodology of flare prediction}
\label{section4}

	Currently most operational flare forecasts are based on expert decision. However, many recent works \citep{Bobra15,Shin16,HadaMuranushi16,Anastasiadis17,Liu17,Raboonik17,Nishizuka16} demonstrated that the Machine-Learning algorithms can be successfully applied for flare prediction. In this Section, we test if it is possible to forecast $\geq{}M\,1.0$ and $\geq{}X\,1.0$ flares, using Machine-Learning algorithms based solely on the LOS magnetic field characteristics. Our approach is to utilize the Support Vector Machine \citep[SVM,][]{Cortes95} classifier for flare forecasting using the Python module ``Scikit-Learn'' \citep{Pedregosa11}. The description of SVM can be found in \citet{Bobra15}, and in Appendix~\ref{AppendixB}.
		
	The computational cost of the SVM classifier scales with the number of cases in the ``train'' data set and the number of features (characteristics, descriptors) as $O(N^{2}\times{}M)$ if $N>>M$. On one hand, a large number of training samples should positively affect the classifier performance. On the other hand, the SVM classifier has many parameters that should be optimized, and the computing time quadratically increases with the size of ``train'' dataset. Thus, any possibility to reduce the number of cases which need to be classified should be utilized. In the previous Section we have found that the flaring cases mostly occur if a specific characteristic exceeds a certain threshold. We have also obtained that the PIL descriptors are more effective in the separation of the positive and negative cases. Thus, we first performed the classification based on the PIL characteristics only. We automatically classified a case as negative if any of its PIL characteristics was below the corresponding threshold. It was found that this procedure allows us to reduce the amount of data for the SVM classification by 74.4\% (leaving about 1/4 of all cases) for the $\geq$M1.0 class flares and by 91.2\% for the $\geq$X1.0 class flares. Only about 11.6\% of positive cases for the $\geq$M1.0 and 14.0\% for the $\geq$X1.0 class flares were misclassified as negative at this stage. To check the validity of this approach, we repeated the training procedure with the threshold values decreased by a factor of two that led to exclusion of 52.2\% of cases (two times more cases need to be classified) for the $\geq$M1.0 class and 72.8\% (three times more cases need to be classified) for the $\geq$X1.0 class cases.	We have also checked how the inclusion of the global AR parameters (AR area, unsigned magnetic flux, maximum strength of magnetic field, and unsigned horizontal gradient) affect the forecasting result by repeating the training procedure with all 11 parameters.
		
	For the SVM training, we normalize the ``train'' dataset following \citet{Nishizuka16}: $Z = (X - \mu{})/\sigma$, where $X$ is a non-normalized data set, $\mu$ is the mean, and $\sigma$ is the standard deviation. We use the same $\mu$ and $\sigma$ parameters to normalize the ``test'' data set. To find the optimal SVM kernel (among the Linear, RBF, Polynomial, and Sigmoid available in the Python Scikit-Learn package) and its parameters, we perform a cross-validation procedure on the ``train'' dataset: divide it into two subsets (one simulating the train data set, and another simulating the test data set) ten times, and then average the SVM results. As a measure of the SVM performance, we use the True Skill Statistics (TSS) metrics defined as:
		
	$$TSS = \frac{TP}{TP+FN} - \frac{FP}{FP+TN},$$		
		
	where $TP$ is the true positive prediction (number of positive cases predicted as positive), $TN$ is the true negative prediction (number of negative cases predicted as negative), $FP$ is the false positive prediction (number of negative cases predicted as positive), $FN$ is the false negative prediction (number of positive cases predicted as negative). The $TSS$ score is not sensitive to the class imbalance ratio (the relative number of positive and negative cases), and is zero for a pure negative prediction (when all cases are predicted as negative). The standard deviation of the $TSS$ was estimated from the scores obtained during the cross-validation procedure with the optimal parameters.

\section{Results}
\label{section5}
	
	In Section~\ref{section3} it was pointed out that the PIL characteristics separate flaring and non-flaring cases more effectively than the global (integrated) characteristics obtained for the whole ARs. The results in Tables~\ref{table2_Mimportance}~and~\ref{table3_Ximportance} demonstrate that all PIL characteristics give approximately the same scores in both tests for both, the $\geq$M1.0 and $\geq$X1.0 flare predictions. Among the global AR characteristics, the highest score is obtained for the unsigned magnetic flux in the AR area, but still it does not exceed the scores for any PIL parameter.
		
	The results of prediction tests based on the PIL parameters only are summarized in the second column of Table~\ref{table4_TSSscores}. For the $\geq$M1.0 class solar flares, we found that the best score of $TSS = 0.76\pm{}0.03$ can be obtained using the ``sigmoid'' SVM kernel (described in Appendix~\ref{AppendixB}) with parameters $C=0.1$, $\gamma{}=0.01$ and $r=0.001$, and the negative/positive class weights of 1/20. Description of these parameters can be found in Appendix~\ref{AppendixB}. The score was derived from the following classification results: $TP = 1932$, $TN = 42382$, $FP = 6654$, $FN = 234$ (including all cases in the test dataset). For the $\geq$X1.0 class solar flares, we obtained $TSS = 0.84\pm{}0.07$ for the same ``sigmoid'' SVM kernel but with different parameters: $C=0.0001$, $\gamma{}=10.0$ and $r=0.0001$, and the negative/positive classes weights of 1/100. This $TSS$ was derived from the following classification results: $TP = 194$, $TN = 44991$, $FP = 6009$, $FN = 8$.
	
	Interestingly, the flare forecasts performed using only the PIL characteristics have almost the same TSS scores as the forecasts based on the full set of characteristics (including both the PIL and global AR characteristics). The TSS scores for the full set of characteristics are summarized in the third column of Table~\ref{table4_TSSscores}. For prediction of $\geq$M1.0 solar flares, the inclusion of global characteristics even decreased the TSS score from $TSS=$0.76 to $TSS=$0.74. For prediction of $\geq$X1.0 flares, we have obtained the same $TSS=$0.84 score.
	
	The last column of Table~\ref{table4_TSSscores} summarizes the results of the classification using the PIL parameters with the pre-classification threshold decreased by a factor of two. The 50\% decrease of the threshold (which results in a smaller number of pre-classified samples) leads to an insignificant increase of TSS for the $\geq$X1.0 flare prediction (from $TSS=$0.84 to $TSS=$0.85) and gives the same TSS for the $\geq$M1.0 flare prediction. Thus, we can conclude that it is possible to pre-classify a significant number of cases (74.4\% for the $\geq$M1.0 class flares and 91.2\% for the $\geq$X1.0 class flares) by applying thresholds to the PIL parameters without a significant decrease of the prediction TSS score.

\section{Discussion and Conclusion}
\label{section6}

	In this paper, we have developed a machine-learning procedure solely based on the line-of-sight (LOS) magnetic field observations that are available in near-real time from space-based and ground-based observatories. The procedure is based on analysis of characteristics of the magnetic field Polarity Inversion Line (PIL) which is automatically identified by performing the magnetogram segmentation formulated as an optimization task. The PIL characteristics were derived from the SDO/HMI magnetograms for each AR with 1\,h cadence. We estimated the effectiveness of these characteristics for forecasting $\geq$M1.0 and $\geq$X1.0 solar flares, and trained the Support Vector Machine (SVM) to maximize the True Skill Statistics (TSS) metrics. Interestingly, the univariate effectiveness scores are similar for all PIL characteristics, probably, because the PIL characteristics (except, possibly, the Maximum gradient across PIL) correlate with each other (depend on the same PIL length or the PIL area that depends on the PIL length).
		
	The obtained True Skill Statistics scores $TSS = 0.76$ for prediction of $\geq$M1.0 class flares, and $TSS = 0.84$ for prediction of $\geq$X1.0 class flares, can be compared with the scores mentioned in other works. For example, \citet{Anastasiadis17} reported $TSS \approx 0.5$ for the prediction of $\geq$C1.0 class flares, \citet{Shin16} received a maximum of $TSS = 0.371$ for $\geq$M1.0 class flares, \cite{HadaMuranushi16}~--- the $TSS = 0.295$ for $\geq$M1.0 class flares, \citet{Liu17}~--- $TSS = 0.50$ for $\geq$M1.0 class flares. On the other hand, our $TSS$ score for $\geq$M1.0 is lower than ones in the works of \citet[][$TSS=$0.817]{Bobra15}, \citet[][$TSS=$0.88 for SVM classifier]{Nishizuka16}, \citet[][$TSS=$0.856]{Raboonik17}. Also, \citet{Nishizuka16} reported a higher TSS score for $\geq$X1.0 class flares ($TSS=0.88$ for SVM classifier). Our results solely based on the line-of-sight magnetic field observations are lower than those obtained with the use of vector magnetograms, but still comparable.
	
	The score for $\geq$M1.0 class flares received in our work is higher than the known expert predictions quoted by \citet{Nishizuka16}: $TSS = 0.50$ for the NICT Space Weather Forecasting Center and $TSS = 0.34$ for the Royal Observatory of Belgium \citep{Devos14}. It is also higher than the $TSS = 0.53$ of the National Oceanic and Atmospheric Administration (NOAA) Space Weather Prediction Center (SWPC) deduced from Table~4 of \citet{Crown12}. For the $\geq$X1.0 flares, again, our result is higher than the expert prediction with $TSS = 0.21$ \citep[the NICT Space Weather Forecasting Center,][]{Nishizuka16} and with $TSS = 0.49$ \citep[SWPC NOAA, deduced from Table 4 of][]{Crown12}. We can conclude that the accurately-tuned machine-learning technique, even if it is solely based on the LOS magnetic field measurements, can compete with the expert-based predictions.
		
	It is necessary to discuss the influence of the data set construction on the prediction results. First, the way of the division of the data set into the ``train'' and ``test'' subsets can change the prediction scores. For example, the shuffled division (when the ``train'' and ``test'' subsets are not consequent in time, but all cases from one AR are kept in one subset) reduces the scores from $TSS = 0.76$ to $TSS = 0.70$ for $\geq$M1.0 class flares, and from $TSS = 0.84$ to $TSS = 0.63$ for $\geq$X1.0 class flares. The strong difference in the TSS score for $\geq$X1.0 class flares is caused by a low number of X-class flares in the data set. In this work, we relied on the NOAA AR detection and considered every case with the detectable PIL, which already makes the data set to be subjective to the PIL detection method. \citet{Nishizuka16} used their own method to detect ARs, which definitely leads to another data set with larger number of cases. \citet{Bobra15} reduced the actual data set by cutting out some randomly-selected portion of negative cases. Thus, to guarantee the accurate comparison of different prediction methods, one should unify the starting data set and its division into the ``train'' and ``test'' subsets. Such attempts were done previously \citep{Barnes16}, and hopefully will be continue in the future.

	The important role of PIL in the flare development process was pointed out in many observations, simulations and forecasts of solar flares. Generally, the PILs are characterized by highly-sheared magnetic fields, strong field gradients and complicated topology of neighboring magnetic field structures. These properties result in a substantial amount of free magnetic energy that can be released in flares. It is not surprising that many flares are developed locally in the PIL vicinity. Our study statistically confirms the importance of the PIL characteristics for flare forecasting. In particular, it demonstrated that the PIL characteristics obtained just from the line-of-sight magnetic field component can be used to obtain flare predictions compatible with expert-based forecasts and comparable to the predictions that are based on full vector magnetic field observations. However, our results are accompanied by a significant number of false positive predictions. Generally, a more accurate comparison of machine-learning-based and expert-based predictions is required. Despite the promising results, we should always keep in mind that the prediction is metrics-dependent. In this work, we maximize the True Skill Statistics in a single parameter setup. Maximizing other metrics can result in other optimal SVM parameters and prediction scores \citep{Bobra15}. Further work is needed to develop algorithms for quantitative prediction of the flare class and physical properties (eruptive or non-eruptive nature, geo-effectiveness etc).

\acknowledgments

Authors thank the anonymous referee for the helpful and detailed review of the paper. Authors thank the GOES and SDO/HMI teams for the availability of high-quality scientific data. Authors also thank D.~Laptev for valuable discussions of the magnetogram segmentation algorithm. The research was partially supported by the NASA Grants NNX14AB68G and NNX16AP05H.

\appendix

\section{Magnetogram Segmentation and PIL Detection Algorithm}
\label{AppendixA}
	
	Suppose $B$ is a magnetic field strength map (magnetogram), $Z_{i}$ is a class of pixel $i$ of the magnetogram (i.e. ``positive'', ``negative'' or ``neutral''), $N$ is the total number of pixels in the magnetogram, $\varepsilon{}(i)$ is a neighborhood (e.g. the closest 8 pixels) of pixel $i$. The magnetogram segmentation can be formulated as the following optimization procedure to maximize function $p(Z,B)$ for a given $B$ by finding optimal classification $Z_{max}$~\citep{Laptev11}:
	$$p(Z_{max},B) = \underset{Z}{max}\ p(Z,B)\propto \prod_{i=1}^{N} \phi{}_{i}(Z_{i},B_{i}) \prod_{j\in{}\varepsilon{}(i)}^{} \phi(Z_{i},Z_{j})$$
	
	Here $\phi{}_{i}(Z_{i},B_{i})$ and $\phi(Z_{i},Z_{j})$ are the scoring functions for each pixel depending on the magnetic field strength and assumed classes of pixels. The choice of the scoring function defines segmentation characteristics and, in fact, should do the following: separate the segments of positive and negative magnetic field polarity, and avoid very small segments with weak field probably coming from noise in the data. We use the scoring functions suggested by \citet{Chernyshov11}:
	\begin{center}
		$ \phi{}_{i}(Z_{i},B_{i}) = e^{-C_{1}\sqrt{|B_{0} - B_{i}|}}$, for $Z_{i}$ ``positive'' \\
		$ \phi{}_{i}(Z_{i},B_{i}) = e^{-C_{1}\sqrt{|B_{0} + B_{i}|}}$, for $Z_{i}$ ``negative'' \\
		$ \phi{}_{i}(Z_{i},B_{i}) = e^{-C_{2}|B_{i}|}$, for $Z_{i}$ ``neutral'' \\
		$ \phi(Z_{i},Z_{j}) = e^{C_{pair}[Z_{i} \ne{} Z_{j}]}, $
	\end{center}
	
	where parameters $C_{1} = 1.0$, $C_{2} = 1.0$, $C_{pair} = 20$, $B_{0} = 1000$\,G are chosen to obtain a stable segmentation of magnetic polarities in strong field regions. Here $[Z_{i} \ne{} Z_{j}]$ is equal 1 if $Z_{i} \ne{} Z_{j}$, and zero otherwise. Following \citet{Laptev11}, the function $p(Z,B)$ is interpreted as conditional probability density function $p(Z|B)$, and is approximated by the factorized probability density function $q(Z) = \prod_{i=1}^{n} q_{i}(Z_{i})$. To measure how strongly the factorized distribution deviates from the actual, one can use the Kullback-Leibler (KL) divergence \citep[][]{Bishop06}. In order to find the best approximating factorized distribution, $q(Z)$, one can minimize the KL divergence:
	
	$$\underset{q(Z)}{min}\ KL(q||p) = -\int q(Z)\textrm{log}\frac{p(Z|B)}{q(Z)}dZ $$
	
	Here we keep the original notation for KL-divergence $KL(q||p)$ between distributions $q$ and $p$ introduced in \citet{Bishop06}. The optimal $q(Z)$ is given by solution of the equation \citep[following][]{Chernyshov11}:
	$$q_{i}(Z_{i}) = \frac{1}{C}\textrm{exp}(\log(\phi{}_{i}(Z_{i})) - C_{pair}\sum_{t\in{} \varepsilon{}(i)}^{} \sum_{j\ne{}i}^{} q_{j}(Z_{j}) ) $$
	
	which can be found iteratively:
	$$q_{i}^{new}(Z_{i}) = \frac{1}{C}\textrm{exp}(\log(\phi{}_{i}(Z_{i})) - C_{pair}\sum_{t\in{} \varepsilon{}(i)}^{} \sum_{j\ne{}i}^{} q_{j}^{old}(Z_{j}) ) $$
	
	Using this equation, one can calculate the factorized distribution multiplier $q_{i}$ for each pixel $i$ and its assumed class $Z_{i}$ (``positive'', ``negative'', or ``neutral''). Because the factorized distribution represents the product of multipliers for each pixel, one can simply maximize $q_{i}(Z_{i})$ for each pixel $i$ separately and obtain $Z_{max}$.
	
	For identification of PIL in active regions, we smooth the original HMI magnetogram using the Gaussian filter with width $\sigma=$1.5$^{\prime{}\prime{}}$, and apply the segmentation algorithm. Then, we apply a morphological dilation procedure separately for positive and negative segments (i.e. expand each segment to include neighboring pixels), and find the PIL as an intersection of the dilated positive and negative segments. Finally, we filter all small islands of the PIL with the number of pixels less than 3\% of the total number of pixels occupied by PIL. This approach is quite robust, and allows us to automatically identify the PIL and calculate magnetic field properties.

\section{Description of the SVM Classifier}
\label{AppendixB}

	The Support Vector Machine \citep[SVM,][]{Cortes95} classifier is the widely-used supervised-learning classification algorithm. The SVM finds a plane in the descriptor space, which optimally separates the positive and negative cases by solving the following functional minimization problem:
	$$ \underset{\omega{},\epsilon{}}{min}\ L = \frac{1}{2}||\omega{}||^{2} + C\sum_{i=1}^{m}W_{i}\epsilon{}_{i},$$
	$$y_{i}(\left<\omega{},x_{i}\right> + b) \geq 1 - \epsilon{}_{i}, \quad \epsilon{}_{i} \geq 0, $$

	where $\omega$ is a vector normal to the separating plane; $i$ is case number in the ``train'' dataset, varying from 0 to $m$; $C$ is a soft margin parameter; $W_{i}$ is the weight of the group which the case $i$ belongs to, $\epsilon{}_{i}$ is a measure of misclassification of case $i$; $y_{i}$ is a constant equal to 1 for positive cases, and -1 for negative cases. After some transformations, this problem becomes a quadratic minimization problem: the functional depends only on scalar products of vectors of characteristics $\left<x_{i},x_{j}\right>$. To achieve better separation between the positive and negative cases, very often the so-called Kernel trick is used. The scalar product of characteristics in the functional is replaced by a function of the characteristics:
	$$ \left<x_{i},x_{j}\right> \rightarrow k(x_{i},x_{j}).$$
	
	In this work, we have tested several kernels available in the Python Scikit-Learn package:
	$$k(x_{i},x_{j}) = \left<x_{i},x_{j}\right>\ (Linear),$$
	$$k(x_{i},x_{j}) = (\gamma{}\left<x_{i},x_{j}\right>)^{d}\ (Polynomial),$$
	$$k(x_{i},x_{j}) = exp(-\gamma{}\left|x_{i}-x_{j}\right|^{2})\ (RBF),$$
	$$k(x_{i},x_{j}) = \tanh(\gamma{}\left<x_{i},x_{j}\right> + r)\ (Sigmoid),$$
	
	where $\gamma$,  $r$ and $d$ are tuning parameters. The other SVM parameters are the soft margin parameter and weights for both classes (multipliers of the soft-margin parameter). One needs to optimize all these parameters during the cross-validation procedure.

\clearpage

\clearpage

\clearpage

\begin{sidewaystable}
	\caption{Relationship between magnetic field characteristics and solar flares of the GOES X-ray classes greater than M1.0 and X1.0. Columns 2 and 3 show the average values of the parameters for the $\geq$M1.0 and $\geq$X1.0 class flares correspondingly. Columns 4 and 5 show the thresholds, above which 95\% of all $\geq$M1.0 and $\geq$X1.0 class flares were observed.}
	\label{table1_parameters}
	\tiny
	\begin{tabular}{cccccc}
		\hline
		Characteristic	&	Average value	&	Average value	&	Average value	&	5\% threshold	&	5\% threshold	\\
			&	(negative cases)	&	(positive $\geq$M1.0 cases)	&	(positive $\geq$X1.0 cases)	&	(positive $\geq$M1.0 cases)	&	(positive $\geq$X1.0 cases) \\
		\hline
		
		PIL length [$m$]	&	(1.7$\pm$1.8)$\cdot$10$^7$	&	(6.8$\pm$4.0)$\cdot$10$^7$	&		(8.3$\pm$3.6)$\cdot$10$^7$	&	1.7$\cdot$10$^7$	&	3.2$\cdot$10$^7$	\\
		
		PIL area [$m^{2}$]	&	(7.3$\pm$5.5)$\cdot$10$^{14}$	&	(19.3$\pm$7.9)$\cdot$10$^{14}$	&	(21.8$\pm$7.6)$\cdot$10$^{14}$	&	7.9$\cdot$10$^{14}$	&	9.6$\cdot$10$^{14}$	\\
		
		Unsigned magnetic flux in the PIL area [$G\cdot{}m^{2}$] &	(1.14$\pm$1.15)$\cdot$10$^{17}$	&	(4.68$\pm$2.98)$\cdot$10$^{17}$	&	(5.96$\pm$2.98)$\cdot$10$^{17}$	&	1.20$\cdot$10$^{17}$	&	1.57$\cdot$10$^{17}$	\\
		
		Unsigned horizontal gradient in the PIL area [$G\cdot{}m$]	&	(0.81$\pm$0.75)$\cdot$10$^{11}$	&	(2.93$\pm$1.67)$\cdot$10$^{11}$	&	(3.40$\pm$1.34)$\cdot$10$^{11}$	&	0.89$\cdot$10$^{11}$	&	1.47$\cdot$10$^{11}$	\\
		
		Maximum gradient across the PIL [$G/m$]	&	(3.8$\pm$2.3)$\cdot$10$^{-4}$	&	(9.0$\pm$4.4)$\cdot$10$^{-4}$	&	(10.3$\pm$3.4)$\cdot$10$^{-4}$	&	3.7$\cdot$10$^{-4}$	&	5.3$\cdot$10$^{-4}$	\\
		
		Gradient-weighted PIL length [$m\cdot{}G/m$]	&	(3.1$\pm$4.1)$\cdot$10$^{3}$	&	(19.4$\pm$16.4)$\cdot$10$^{3}$	&	(24.1$\pm$13.2)$\cdot$10$^{3}$	&	2.8$\cdot$10$^{3}$	&	6.2$\cdot$10$^{3}$	\\
		
		R-value [$G\cdot{}m^{2}$]	&	(2.4$\pm$3.2)$\cdot$10$^{15}$	&	(14.2$\pm$11.7)$\cdot$10$^{15}$	&	(19.1$\pm$10.7)$\cdot$10$^{15}$	&	2.0$\cdot$10$^{15}$	&	4.8$\cdot$10$^{15}$	\\
		
		AR area [$m^{2}$]	&	(4.8$\pm$4.0)$\cdot$10$^{15}$	&	(10.1$\pm$4.9)$\cdot$10$^{15}$	&	(11.9$\pm$4.7)$\cdot$10$^{15}$	&	3.2$\cdot$10$^{15}$	&	3.7$\cdot$10$^{15}$	\\
		
		Unsigned magnetic flux in the AR area [$G\cdot{}m^{2}$]	&	(7.7$\pm$7.1)$\cdot$10$^{17}$	&	(21.1$\pm$13.0)$\cdot$10$^{17}$	&	(29.2$\pm$13.0)$\cdot$10$^{17}$	&	5.6$\cdot$10$^{17}$	&	7.1$\cdot$10$^{17}$	\\
		
		Maximum strength of magnetic field in the AR [$G$]	&	(1.31$\pm$0.41)$\cdot$10$^{3}$	&	(1.66$\pm$0.48)$\cdot$10$^{3}$	&	(1.84$\pm$0.52)$\cdot$10$^{3}$	&	1.06$\cdot$10$^{3}$	&	1.20$\cdot$10$^{3}$	\\
		
		Unsigned horizontal gradient in the AR area [$G\cdot{}m$]	&	(6.1$\pm$5.4)$\cdot$10$^{11}$	&	(13.4$\pm$7.4)$\cdot$10$^{11}$	&	(16.5$\pm$8.2)$\cdot$10$^{11}$	&	3.9$\cdot$10$^{11}$	&	4.3$\cdot$10$^{11}$	\\
		
		\hline
	\end{tabular}
\end{sidewaystable}
\clearpage

\begin{table}
	\caption{Importance of magnetic field characteristics for the forecast of $\geq$M1.0 class solar flares.}
	\label{table2_Mimportance}
	\footnotesize
	\begin{tabular}{ccc}
		\hline
		Characteristic	&	Fraction of negative	&	F-score	\\
			&	cases below threshold, \%	&		\\
		\hline
		PIL length (log)	&	0.63 	&	1.41	\\
		PIL area	&	0.60	&	1.46	\\
		Unsigned magnetic flux in the PIL area (log)	&	0.63	&	1.41	\\
		Unsigned horizontal gradient in the PIL area (log)	&	0.64	&	1.48	\\
		Maximum gradient across the PIL (log)	&	0.56	&	1.15	\\
		Gradient-weighted PIL length (log)	&	0.62	&	1.45	\\
		R-value (log)	&	0.61	&	1.35	\\
		AR area (log)	&	0.44	&	0.66	\\
		Unsigned magnetic flux in the AR area (log)	&	0.49	&	0.86	\\
		Maximum strength of magnetic field in the AR (log)	&	0.29	&	0.30	\\
		Unsigned horizontal gradient in the AR area	&	0.44	&	0.69	\\
		\hline
	\end{tabular}
\end{table}
\clearpage

\begin{table}
	\caption{Importance of magnetic field characteristics for the forecast of $\geq$X1.0 class solar flares.}
	\label{table3_Ximportance}
	\footnotesize
	\begin{tabular}{ccc}
		\hline
		Characteristic	&	Fraction of negative	&	F-score	\\
		&	cases below threshold, \%	&		\\
		\hline
		PIL length	&	0.84	&	2.68	\\
		PIL area	&	0.71	&	2.36	\\
		Unsigned magnetic flux in the PIL area	&	0.74	&	2.51	\\
		Unsigned horizontal gradient in the PIL area	&	0.83	&	2.81	\\
		Maximum gradient across the PIL	&	0.79	&	2.46	\\
		Gradient-weighted PIL length (log)	&	0.84	&	2.62	\\
		R-value (log)	&	0.84	&	2.47  	\\
		Total AR area	&	0.51	&	1.32  	\\
		Unsigned magnetic flux in the AR area (log)	&	0.60	&	1.91	\\
		Maximum strength of magnetic field in the AR (log)	&	0.41	&	0.68	\\
		Unsigned horizontal gradient in the AR area (log)	&	0.49	&	1.29	\\
		\hline
	\end{tabular}
\end{table}
\clearpage

\begin{table}
	\caption{Comparison of TSS scores for different methods of prediction of $\geq$M1.0 and $\geq$X1.0 class solar flares. The standard deviations are estimated using a cross-validation procedure.}
	\label{table4_TSSscores}
	\footnotesize
	\begin{tabular}{|c|c|c|c|}
		\hline
			&	PIL characteristics only	&	PIL + global characteristics	&	50\% decreased cutoff values \\
		\hline
		Prediction of $\geq$M1.0 flares	&	0.76$\pm$0.03	&	0.74$\pm$0.03	&	0.76$\pm$0.03	\\
		\hline
		Prediction of $\geq$X1.0 flares	&	0.84$\pm$0.07	&	0.84$\pm$0.07	&	0.85$\pm$0.04	\\
		\hline
	\end{tabular}
\end{table}
\clearpage

\end{document}